\documentclass[prl,twocolumn,superscriptaddress,noshowpacs]{revtex4}

\usepackage{graphicx}
\usepackage{dcolumn}
\usepackage{bm}

\usepackage{amsmath}
\usepackage{graphicx}
\usepackage{color}
\usepackage{color}
\usepackage{amssymb}
\usepackage{amsfonts}

\usepackage[colorlinks,urlcolor=blue,linkcolor=blue,citecolor=blue,anchorcolor=blue]{hyperref}

\begin{document}
\title{Particle-like behavior of topological defects in linear wave packets in photonic graphene}
\author{Zhaoyang Zhang}
\affiliation{Key Laboratory for Physical Electronics and Devices of the Ministry of Education \& Shaanxi Key Lab of Information Photonic Technique, School of Electronic and Information Engineering, Xi'an Jiaotong University, Xi'an 710049, China}
\author{Feng Li}
\email{felix831204@xjtu.edu.cn}
\affiliation{Key Laboratory for Physical Electronics and Devices of the Ministry of Education \& Shaanxi Key Lab of Information Photonic Technique, School of Electronic and Information Engineering, Xi'an Jiaotong University, Xi'an 710049, China}
\affiliation{Department of Physics and Astronomy, University of Sheffield, Sheffield, S3 7RH, UK}
\author{G. Malpuech}
\affiliation{Institut Pascal, PHOTON-N2, University Clermont Auvergne, CNRS, 4 avenue Blaise Pascal, 63178 Aubi\`{e}re Cedex, France}
\author{Yiqi Zhang}
\affiliation{Key Laboratory for Physical Electronics and Devices of the Ministry of Education \& Shaanxi Key Lab of Information Photonic Technique, School of Electronic and Information Engineering, Xi'an Jiaotong University, Xi'an 710049, China}
\author{O. Bleu}
\affiliation{Institut Pascal, PHOTON-N2, University Clermont Auvergne, CNRS, 4 avenue Blaise Pascal, 63178 Aubi\`{e}re Cedex, France}
\author{S. Koniakhin}
\affiliation{Institut Pascal, PHOTON-N2, University Clermont Auvergne, CNRS, 4 avenue Blaise Pascal, 63178 Aubi\`{e}re Cedex, France}
\author{Changbiao Li}
\affiliation{Key Laboratory for Physical Electronics and Devices of the Ministry of Education \& Shaanxi Key Lab of Information Photonic Technique, School of Electronic and Information Engineering, Xi'an Jiaotong University, Xi'an 710049, China}
\author{Yanpeng Zhang}
\email{ypzhang@xjtu.edu.cn}
\affiliation{Key Laboratory for Physical Electronics and Devices of the Ministry of Education \& Shaanxi Key Lab of Information Photonic Technique, School of Electronic and Information Engineering, Xi'an Jiaotong University, Xi'an 710049, China}
\author{Min Xiao}
\email{mxiao@uark.edu}
\affiliation{Department of Physics, University of Arkansas, Fayetteville, Arkansas, 72701, USA}
\affiliation{National Laboratory of Solid State Microstructures and School of Physics, Nanjing University, Nanjing 210093, China}
\author{D. D. Solnyshkov}
\affiliation{Institut Pascal, PHOTON-N2, University Clermont Auvergne, CNRS, 4 avenue Blaise Pascal, 63178 Aubi\`{e}re Cedex, France}

\date{\today}

\begin{abstract}
\noindent
Topological defects, such as quantum vortices, determine the properties of quantum fluids. Their study has been at the center of activity in solid state and BEC communities. On the other hand, the non-trivial behavior of wavepackets, such as the self-accelerating Airy beams, has also been intriguing physicists. Here, we study the formation, evolution, and interaction of optical vortices in wavepackets at the Dirac point in photonic graphene. We show that while their exact behavior goes beyond the Dirac equation and requires a full account of the lattice properties, it can be still approximately described by an effective theory considering the phase singularities as "particles". These particles are capable of mutual interaction, with their trajectory obeying the laws of dynamics.
\end{abstract}

\maketitle

Topological invariants \cite{Nobel2016,Thouless1998} become as important in physics as the symmetries \cite{Gross1996}. They open a new dimension for the exploration of fundamental possibilities and the creativity of engineering. Quantum Hall effect \cite{Klitzing1980,TKNN1982,Kohmoto1984,Hatsugai1993} and topological insulators \cite{Kane2005,Kane2005a,Hasan2010} have shown that the band structure of periodic systems is not limited to mere dispersion, and that there can be chiral edge states  protected by topological invariants of the bands. The topological invariants can characterize not only such rigid structures as the bands, but also define the properties of the quantum fluids by determining the existence of topological defects \cite{Kosterlitz1972,Berezinskii1972}. Indeed, quantum vortices, discovered in superconductors \cite{Abrikosov1957}, liquid helium \cite{Onsager1949,Feynman1956}, and Bose condensates \cite{Ketterle2001}, are protected by a topological invariant -- their winding. While being different from the famous Chern number \cite{Chern1946}, characterizing the bands, the winding number \cite{Neill1967,Thouless1998}, defined in differential geometry, is also a well-known topological invariant in mathematics, providing associated protection.

Fluids are usually interacting \cite{LandauFM}, and quantum vortices have mostly been analyzed in interacting systems: for example, in superconductors and in Bose condensates, the vortex size is determined by the interactions \cite{LeggettBook}. However, topological defects (called phase singularities or dislocations in this case) can also be observed in linear (non-interacting) wave interference \cite{Berry1974}, not only for light \cite{Braunbek1952}, but also for tidal waves \cite{Whewell1833,Berry2000}.
Phase singularities, being just zero density points, are less limited by physical bounds: for example, their speed can exceed the speed of light \cite{Berry1974}, and there is in general no strict connection between the number of vortices and the optical angular momentum of a beam \cite{Berry2009}, except for simple cases (Gauss-Laguerre beams). Because of this, topological defects in wave interference were seen as being objects somewhat "less real" than similar defects in the interacting quantum fluids. The distinction between the interacting and non-interacting case has become a matter of debate \cite{Cilibrizzi2014,Comment2015}, because in many works vortices were used as a smoking gun of superfluidity \cite{Zwierlein2005,Lagoudakis2008}.

Wavepackets in linear regime are an important field on their own. The famous self-accelerated Airy beams \cite{Berry1979,Siviloglou2007} are one example, but there are also Bessel beams \cite{Durnin1987}, with their self-repairing properties \cite{Bouchal1998} and spatial profile, which are particularly useful for applications such as optical tweezers \cite{Garces2002}. Even the physics of Gaussian wavepackets in non-trivial systems with diabolical points, such as honeycomb lattices, has been attracting attention since a very long time, with original phenomena such as the conical refraction, predicted \cite{Hamilton1837} and observed \cite{Lloyd1833} a long time ago. A finalized theory describing the intensity evolution in such wavepackets was developed only recently \cite{Belskii1978,Berry2006}. The phase properties of conical diffraction are understood even less.  Recently, the conversion of pseudospin into orbital angular momentum has been described for such wavepackets at the Dirac point \cite{Song2015}: a vortex has been shown to appear in the center of the wavepacket after its evolution in the effective field of the Dirac Hamiltonian. Another work has shown the formation of several vortices in photonic Lieb lattices \cite{Diebel2016}, but their dynamical behavior has not been analyzed.

In this work, we show that optical vortices appearing in linear wavepackets exhibit many features typical for topological defects in nonlinear quantum fluids. Their trajectories obey the laws of dynamics: in particular, we observe the effect of the Magnus force and the mutual interaction of two vortices. Finally, we show that certain features of wavepackets even in the immediate vicinity of the Dirac point in graphene cannot be described by the Dirac equation, because the two pseudospin components actually coexist in the same real space.

We study the evolution of a probe beam in a honeycomb lattice (photonic graphene). The transverse beam profile can be found by looking for the solution of the wave equation in the form $\vec{E}(x,y,z,t)=\vec{E_0}a(x,y,z)e^{i(k_0nz-\omega t)}$, where $\omega$ is the frequency of the laser beam, $n$ is the refraction coefficient, $k_0$ is the wave vector of light in the vacuum, $\vec{E_0}$ is the maximal amplitude vector, and $a$ determines the spatial intensity distribution. The paraxial approximation $\partial^2 a/\partial z^2\ll k_0\partial a/\partial z$ allows rewriting the equation as:
\begin{equation}
\label{parax}
  i\frac{{\partial a}}{{\partial z}} =  - \frac{1}{{2{k_0}{n_0}}}\Delta a - k_0^2\left( {{n^2} - n_0^2} \right)a,
\end{equation}
which is equivalent to the Schr\"odinger equation, where the propagation in the $z$ direction is mapped to time $t$, the mass is determined by $m=\hbar k_0 n_0/c$ ($n_0$ is the background refraction index, $c$ is the speed of light), and the potential $U(x,y)$ is determined by the deviation of the refraction index from the background value: $U(x,y)=-\hbar c k_0^2(n^2-n_0^2)$. In the vicinity of the Dirac point of the honeycomb lattice the behavior of the wavepackets is supposed to obey the Dirac equation:
\begin{equation}
i\hbar\frac{\partial\psi}{\partial t}=\hbar c' \mathbf{k}\cdot\mathbf{\sigma} \psi
\label{dirac1}
\end{equation}
where $\psi=(\psi_A,\psi_B)^T$ is a spinor wavefunction with two components (in the case of graphene, these are the wavefunctions on the two sites of the unit cell $A$ and $B$), and $c'$ is the effective speed of light determined by the microscopic Hamiltonian (e.g. Fermi velocity). Note that the full solution $a(x,y,z)$ of Eq.~\eqref{parax} also includes a plane wave of the $K$ point \cite{suppl}.

\begin{figure}[tbp]
\includegraphics[width=1\linewidth]{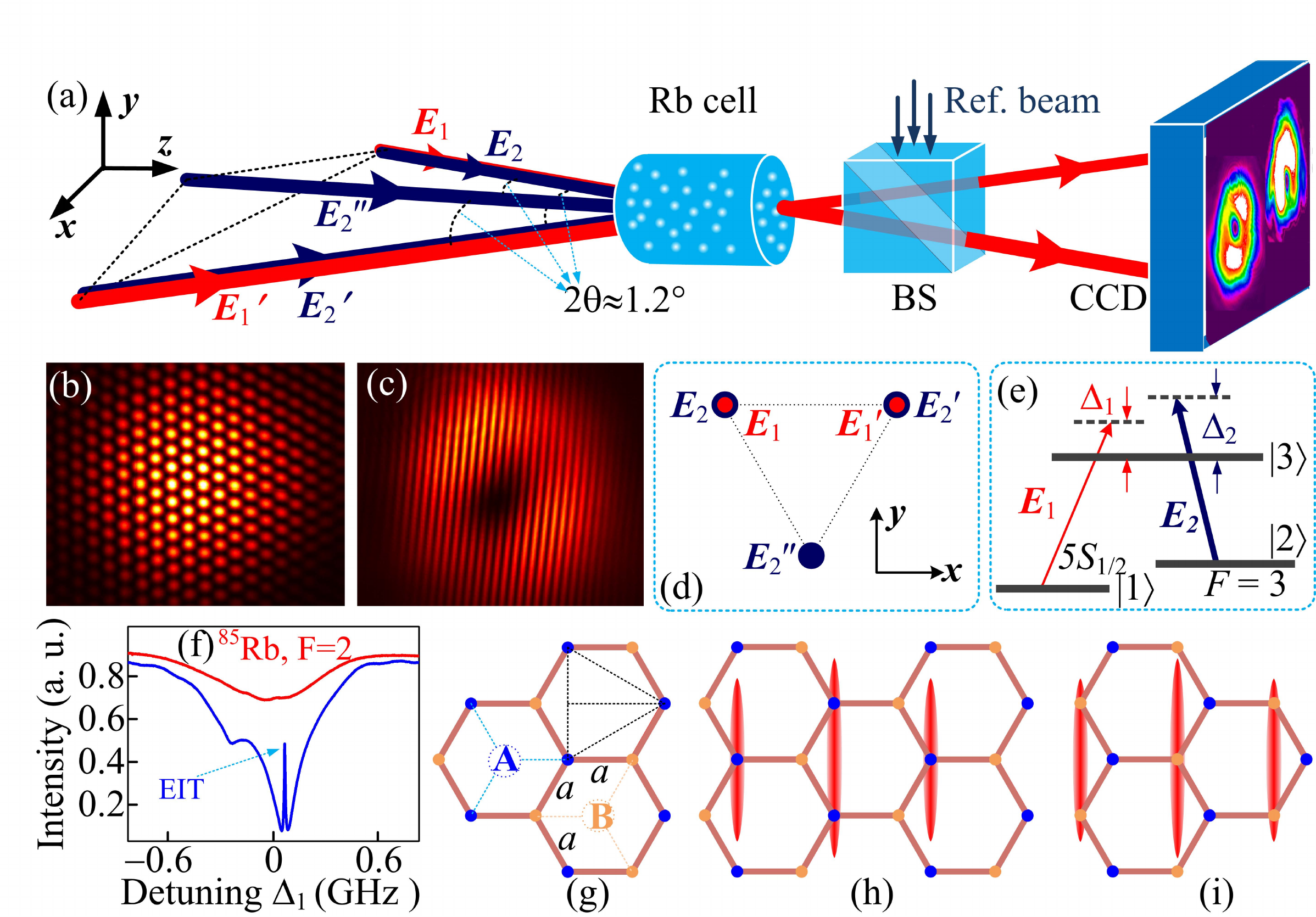}
\caption{\label{figExp1} (Color online) (a) Experimental setup. $\emph{\textbf{E}}_{2}$, $\emph{\textbf{E}}_{2}'$ and $\emph{\textbf{E}}_{2}''$are three coupling beams interfering to form a hexagonal optical lattice inside the rubidium vapor cell, which results in a honeycomb lattice for susceptibility due to the EIT effect.  $\emph{\textbf{E}}_{1}$ and $\emph{\textbf{E}}_{1}'$ are two probe beams from the same laser that construct a probe field featured by equally spaced vertical fringes. BS: 50/50 beam splitter, CCD: charge coupled device camera. (b) The observed hexagonal coupling lattice. (c) The interference fringes formed by the two probe beams, both of which are set to carry an OAM=1. (d) Spatial beam arrangement (before entering the cell) of the probe and coupling beams in the \emph{x-y} plane. (e) The three-level $\Lambda$-type energy-level structure coupled by the probe and coupling fields. (f) The observed absorption (upper red curve, corresponding to the transition $^{85}$Rb, $F=2$ $\rightarrow$ $F'$) and EIT spectra (lower blue curve) versus the probe-field detuning. (g) The schematic for the generated hexagonal lattice. (h) and (i) The beam arrangements of the periodic probe field and the induced honeycomb lattice inside the medium for exciting A and B sublattices, respectively.}
\end{figure}

The photonic graphene is formed in a $^{85}$Rb vapor cell by electromagnetically induced transparency(EIT)\cite{PhysRevA.51.576}, as illustrated in Fig. \ref{figExp1}. Generally, the susceptibility experienced by a probe field $\emph{\textbf{E}}_{1}$ in the $\Lambda$-type three-level $^{85}$Rb atomic configuration [Fig. \ref{figExp1}(e)] under the effect of a coupling field $\emph{\textbf{E}}_{2}$ reads \cite{PhysRevA.51.R2703,PhysRevA.97.013603}
\begin{equation}
\chi  = \frac{{iN{{\left| {{\mu _{31}}} \right|}^2}}}{{\hbar {\varepsilon _0}}} \times \frac{1}{{\left( {{\Gamma _{31}} + i{\Delta _1}} \right) + \frac{{{{\left| {{\Omega _2}} \right|}^2}}}{{{\Gamma _{32}} + i\left( {{\Delta _1} - {\Delta _2}} \right)}}}}
\label{sus}
\end{equation}
where $\varepsilon_{0}$ is the vacuum dielectric constant; $\Gamma_{31}$ (resp. $\Gamma_{32}$) is the decay rate between states $|1\rangle$ (resp. $|2\rangle$) and $|3\rangle$; $N$ is the atomic density at  $|1\rangle$. $\Delta_{1}$ (resp. $\Delta_{2}$) is the frequency detuning between the atomic resonance $|1\rangle$ to $|3\rangle$(resp. $|2\rangle$ to $|3\rangle$) and the probe (resp. coupling) field frequency, as labeled in Fig. \ref{figExp1}(e). $\Omega_{2}$ is the Rabi frequency induced by the coupling field $\emph{\textbf{E}}_{2}$ and $\mu_{31}$ is the dipole moment between levels $|1\rangle$ and $|3\rangle$. The coupling field is constructed by the interference of 3 laser beams which induces a honeycomb-like susceptibility distribution \cite{zhangyiqi.lpr.9.331.2015} with a negligibly small imaginary part \cite{suppl}. The probe field is also structured to form periodical vertical fringes by 2-beam interference \cite{suppl}, to allow selective coverage of only one set (either A or B) of the sublattices [Fig.~\ref{figExp1}(c), (g)-(i)], and to excite the $K$ or $K'$ valley in the momentum space \cite{Song2015}. In our experiment, the probe can be either Gaussian or Gauss-Laguerre with an orbital angular momentum (OAM). After exiting the Rb cell, the two probe beams that construct the probe field separate in space [Fig. \ref{figExp1}(a)], and interfere with a Gaussian reference beam. We record such interference pattern for one of the two probe beams on a charge coupled device (CCD) camera.

\begin{figure}[tbp]
\includegraphics[width=0.9\linewidth]{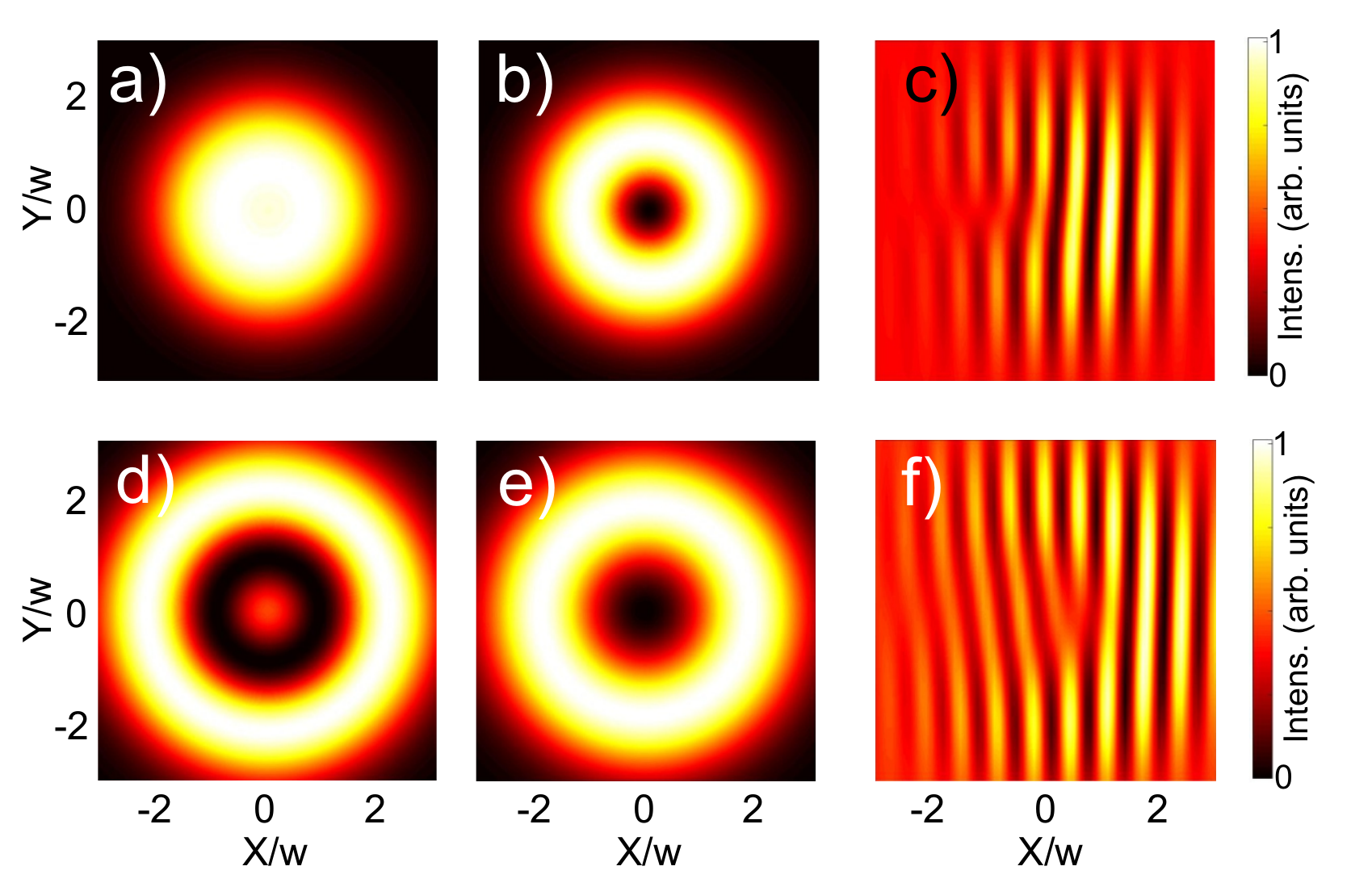}
\caption{\label{fig2} (Color online) Wavepacket in the Dirac equation. The 2 rows correspond to $t/T=0.67~(1.17)$. Intensity: (a,d) $|\psi_A|^2$, (b,e) $|\psi_B|^2$. Interference: (c,f) $|\psi_A+\psi_B+e^{i\mathbf{k_rr}}|^2$, $k_rw=15$.}
\end{figure}

\begin{figure}[tbp]
\includegraphics[width=0.95\linewidth]{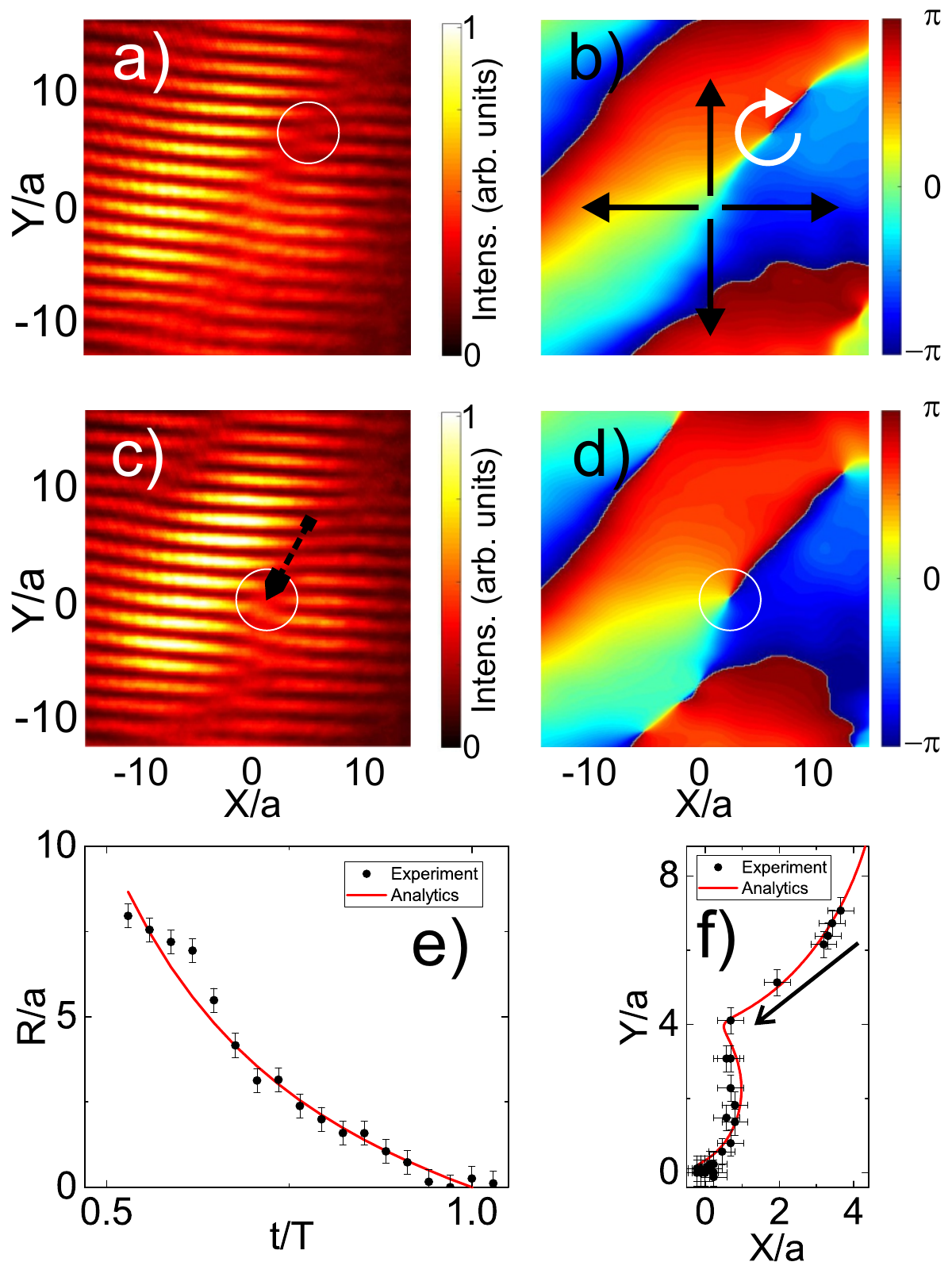}
\caption{\label{fig2B} (Color online) Experimental wavepacket evolution. The 1st two rows correspond to $\Delta_1=45~(56)$~MHz. (a,c) the interference pattern with a reference beam; (b,d)  extracted phase [arrows in panel (b) show different contributions to the current]. The last row  shows the vortex trajectory (e)
$r(t)$ and (f) $Y(X)$ (black dots - experiment, red line - analytical theory).}
\end{figure}

The massless Dirac equation \eqref{dirac1}  is characterized by a single parameter $c'$. Therefore, changing this parameter is equivalent to changing the units of time. In the experiment, varying the detuning $\Delta_1$ changes the susceptibility of the honeycomb-like atomic superlattice [Eq. \eqref{sus}], which changes the amplitude of the potential and the tunneling probability between the lattice sites, and thus the effective units of time (see \cite{suppl}).

The non-trivial behavior of wavepackets in the Dirac equation has been in the focus of theoretical studies for a long time \cite{Eckart1936}, and several corresponding experiments have also appeared recently \cite{Song2015}. Different representations (centered on Berry curvature \cite{Chang2008,Xiao2010} or on the effective field \cite{Song2015}) provide different levels of comprehension, but the conclusion is generally the same: wavepackets in the Dirac equation are almost always associated with a nonzero angular momentum \cite{Chang2008}, which can be seen (and observed experimentally) as a quantum vortex.

We begin with a simple case with zero angular momentum of the initial wavefunction $\psi_A=\exp(-r^2/2w^2)$. The Hamiltonian \eqref{dirac1} converts $\psi_A$ to $\psi_B$, but because of its dependence on the polar angle of $\mathbf{k}$, the resulting conical refraction is accompanied by the change of winding: $l_B=l_A\pm1$.  Figure~\ref{fig2} shows the calculated images of the evolution of a Gaussian wave packet in the Dirac equation, exhibiting conical refraction: $|\psi_A|^2$ in panels (a,d), $|\psi_B|^2$ in (b,e), and the interference of their superposition with a reference beam in (c,f). The phases $\arg(\psi_A)$ and $\arg(\psi_B)$ do not change with time: the phase singularity is always present only in the center of $\psi_B$.  For times shorter than the period $T$ of component conversion $t<T=w/c$, the approximate solution along $x$ reads:
\begin{equation}
\begin{gathered}
  {\psi _A}\left( {x,t} \right) = A\left( t \right)\exp \left( { - \frac{{{x^2}}}{{2{w^2}}}} \right) \hfill \\
  {\psi _B}\left( {x,t} \right) =  B\left( t \right)\frac{xct}{{{w^2}}}\exp \left( { - \frac{{{x^2}}}{{2{w^2}}}} \right) \hfill \\
\end{gathered}
\end{equation}
where: $A(t)=\cos\omega t$ and $B(t)=\sin\omega t$ ($\omega=2\pi/T$). 

In experiment, the emission is detected far from the Rubidium cell, and individual sites cannot be distinguished. Therefore, the total emission detected is a \emph{superposition} of $\psi_A$ and $\psi_B$, with the phase being that of a superposition of two complex fields. The motion of a vortex in the superposition $\psi_A+\psi_B$ is a result of the interplay of the intensities of the components, and its position can be found from the simple equation $\psi_A+\psi_B=0$. This can only occur along the $x$ axis, because the phase of $\psi_A$ and $\psi_B$ is opposite for negative $x$. The equation for the vortex position reads $A\left( t \right) + B\left( t \right)xct/w^2 = 0$
which  allows to find the trajectory $x(t)$:
\begin{equation}
\left|x\left( t \right)\right| = \frac{{{w^2}\cot \omega t}}{{ct}}
\label{traj}
\end{equation}

Figure~\ref{fig2B}(a,c) shows the experimental images of the interference of the transmitted beam with a reference beam at 2 detunings corresponding to 2 different times (see also movies in \cite{suppl}). The vortex position is visible as a forklike dislocation (white circle), and its shift is marked by a black arrow. Panels (b,d) show the extracted phase. The extracted vortex trajectory (with the error bars determined by the interference fringes) is compared in Fig.~\ref{fig2B}(e) with the analytical solution Eq.~\eqref{traj} (red curve). The good quality of the fit confirms the interpretation (the origin of the experimental time axis is the fitting parameter). In the Dirac equation, vortex appears at infinity and approaches the system center very rapidly in the initial moments. In the experiment, it appears as a part of a vortex-antivortex pair at a finite distance, determined by the sensitivity of the detector and the finite size of the photonic graphene lattice (see \cite{suppl}).

However, the exact cycloidal experimental XY trajectory of the vortex [Fig.~\ref{fig2B}(f), black dots] cannot be simulated with the Dirac equation, but can be reproduced only if one takes into account the Magnus force \cite{Ao1993,Thouless1996}
\begin{equation}
\mathbf{F}=I\hbar(\mathbf{L}\times\mathbf{v})
\end{equation}
where $I$ is the relative intensity (fitting parameter), $\mathbf{L}$ is the vortex winding and $\mathbf{v}$ is the vortex velocity obtained from Eq.~\eqref{traj}. The Magnus force was used to detect a single vortex in superfluid helium for the first time \cite{Vinen1958}, here we use it to prove the "reality" of the phase singularity and of the associated rotation in a two-component light beam [Fig.~\ref{fig2B}(f), red curve]. While in the Dirac approximation, the phase singularity is present only in the center of the wavepacket in the $\psi_B$ component, in experiment and in full numerical simulations it is located at the center of the vortex, as can be seen from the experimental phase images [Fig.~\ref{fig2B}(b,d)]. Therefore, the overall microscopic outward flow due to the intensity gradient [Fig.~\ref{fig2B}(b), black arrows]  is increased on one side of the singularity because of the extra flow of the vortex (white arrow), which accelerates the equilibration of the population of A and B sites and shifts the vortex (given by $\psi_A=-\psi_B$) laterally. The mechanism is therefore the same as that of the common Magnus force in classical and quantum fluids. In spite of the fact that the phase singularity itself is a zero-density point, not restricted by the laws of relativity, its motion is still affected by the flow it creates.

\begin{figure}[tbp]
\includegraphics[width=1\linewidth]{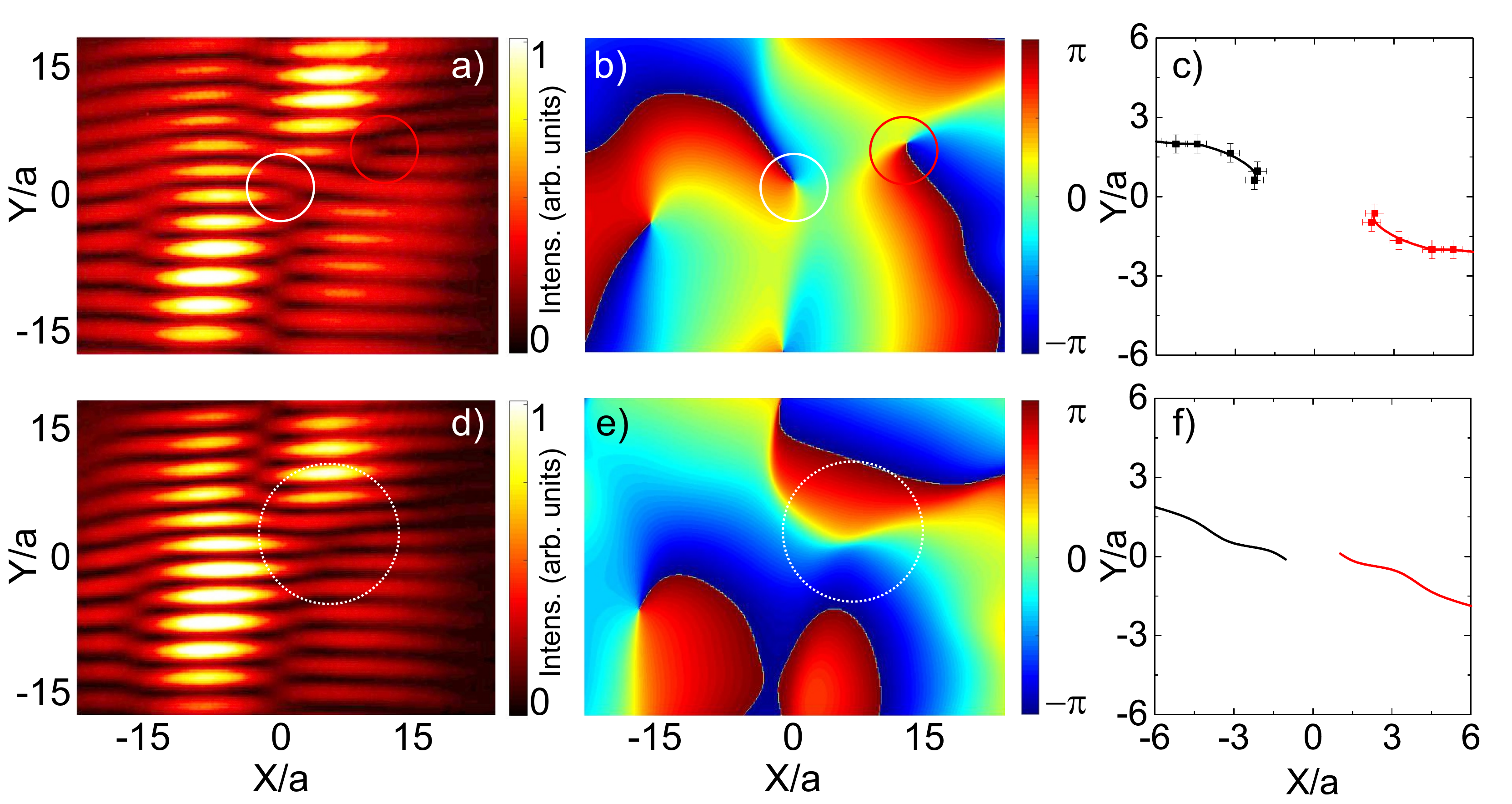}
\caption{\label{fig3} (Color online) Experiment and theory for initial wavepacket with $L=+1$. The two rows correspond to $\Delta_1=80.4~(96.4)$~MHz: (a,d) - interference pattern (experiment), (b,e) - extracted phase.
Vortex trajectories: (c) - experiment (curves are guides for the eyes), (f) - theory.}
\end{figure}

We have also studied the evolution of wavepackets with non-zero OAM. The conversion of the component $\psi_A$ into $\psi_B$ by the Dirac Hamiltonian \eqref{dirac1} changes the angular momentum by $1$: $l_B=l_A-1$. If the injected wavepacket has a positive angular momentum $l_A=1$, then the other component has $l_B=0$, and the overall interference pattern shows a vortex and an anti-vortex, canceling each other at $t=T$. In the other case, when $l_A=-1$, $l_B=-2$, and the overall pattern shows $L=-2$ at $T$.
Here, we use the full numerical treatment in the paraxial approximation, beyond the Dirac approximation.
Figure~\ref{fig3} shows the experimental and theoretical images for a wavepacket with $L=+1$. Panels (a,d) show the experimental interference patterns, with vortices marked by ellipses. In panel (d), two opposite vortices meet and disappear (dashed ellipse). This is confirmed by the phase images (b - two dislocations, e - no dislocation). The trajectories of the two vortices in panel (c) are well reproduced by theoretical modeling (f) (see also \cite{suppl}) [fitting parameter -- initial position of the vortex $(-0.2,-0.5)a$].

\begin{figure}[tbp]
\includegraphics[width=0.95\linewidth]{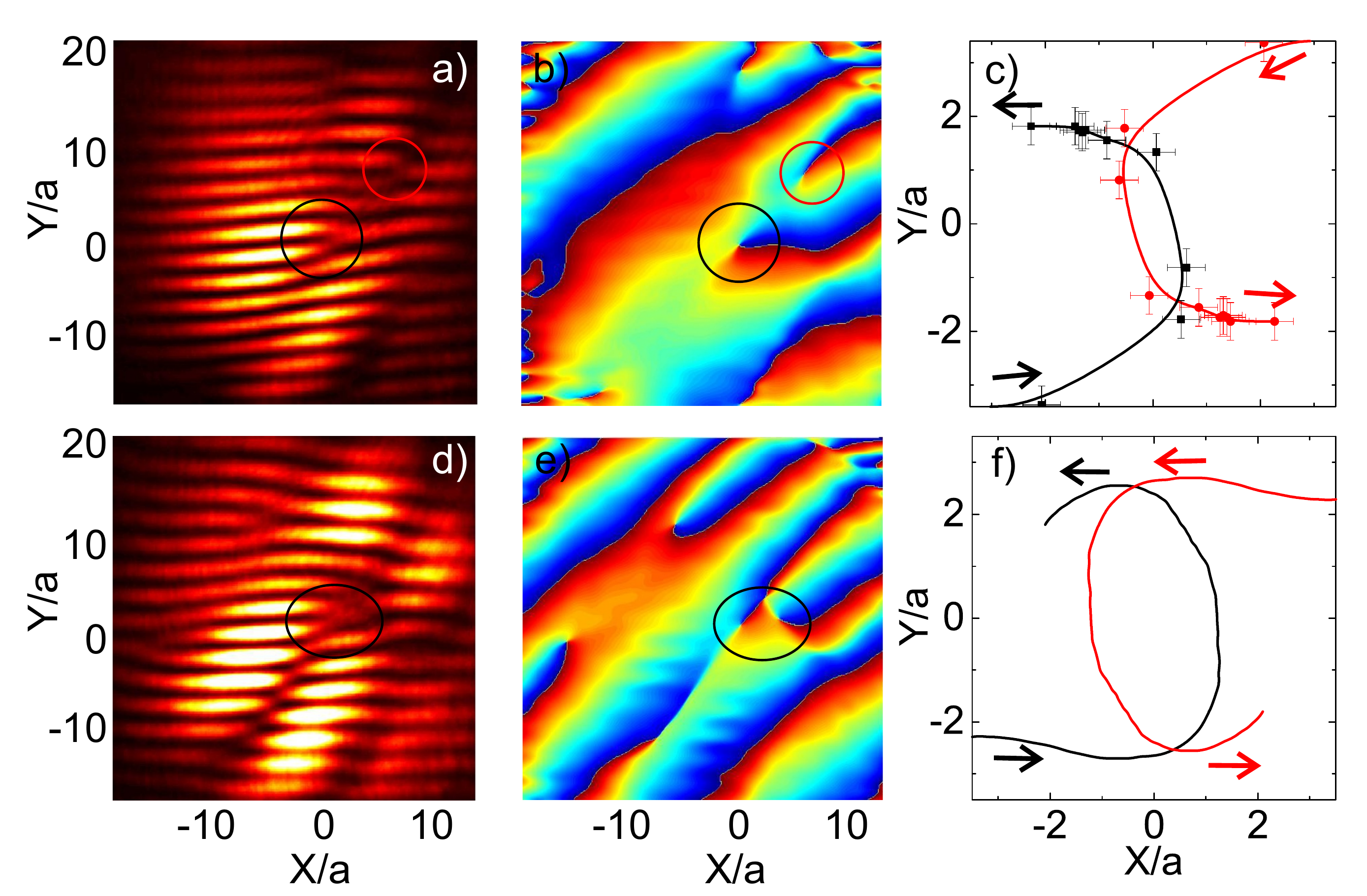}
\caption{\label{fig4} (Color online) Experiment and theory for initial wavepacket with $L=-1$. The two rows correspond to $\Delta_1=50~(59.6)$~MHz. (a,d) - experimental interference patterns (experiment), (b,e) - extracted phase. Trajectories of the vortices: (c) experiment (smooth curves are guides for the eyes), (f) theory.}
\end{figure}

Figure~\ref{fig4} demonstrates the evolution of a wavepacket with $L=-1$ (movie in \cite{suppl}). In this case, two vortices of the same sign appear in the experimental interference patterns (a,d) and phase (b,e). As in the case $L=0$, it is possible to find an analytical solution for the vortex trajectory from the condition $\psi_A=-\psi_B$ (see \cite{suppl}), but it is again a straight line. However, in reality, each of the 2 vortices creates a velocity field which affects the other, leading to their mutual rotation around their center of mass, as can be seen from the experiment (c) and full numerical simulations (f) [fitting parameter -- initial position of the vortex $(-0.46,-1.8)a$].

Neither the cycloidal motion due to the Magnus force visible in Fig.~\ref{fig2B}(f), nor the mutual rotation seen in Fig.~\ref{fig4}(c, f) can be reproduced with the Dirac Hamiltonian \eqref{dirac1}, because it neglects the fact that the two components $\psi_A$ and $\psi_B$ actually coexist in the same space, occupying different points. The component conversion $\psi_A\to \psi_B$ thus already corresponds to motion (a flow of intensity), which is neglected. This is a fundamental limitation of the Dirac equation restricting its validity. Indeed, while the wave packet as a whole involves wavevectors close to the Dirac point, determining the position of the vortex core with a high precision involves wavevectors much further from this point, due to the Heisenberg uncertainty principle. Both of these effects stem from the current induced by the phase gradient associated with vortices,  present both in linear and nonlinear systems. The advantage of our configuration is that it allowed us to evidence the consequences of this phase gradient for the phase singularity itself.
The study of the dynamics of vortices is important for future applications, such as vortex memories \cite{Sigurdsson2014,Lorenzo2018} and gyroscopes \cite{Schwab1997,Eckel2014}.

To conclude, we have studied both experimentally and theoretically the behavior of OAM wavepackets in photonic graphene, showing that phase singularities in linear wavepackets can behave as vortices in quantum fluids, exhibiting the effects of the Magnus force and demonstrating mutual interaction. We also point out the limitations of the Dirac equation for the description of the systems with pseudospin defined in real space.

\begin{acknowledgments}
This work was supported by National Key R\&D Program of China (2018YFA0307500,2017YFA0303703), National Natural Science Foundation of China (61605154, 11474228), Natural Science Foundation of Shaanxi Province (2017JQ6039, 2017JZ019 and 2016JM6029), China Postdoctoral Science Foundation (2016M600776, 2016M600777, and 2017T100734) and Postdoctoral Science Foundation of Shaanxi Province (2017BSHYDZZ54 and 2017BSHTDZZ18). We acknowledge the support of the project "Quantum Fluids of Light"  (ANR-16-CE30-0021), of the ANR Labex Ganex (ANR-11-LABX-0014), and of the ANR Labex IMobS3 (ANR-10-LABX-16-01), of the I-Site "Cap2025". D.D.S. acknowledges the support of IUF (Institut Universitaire de France)
\end{acknowledgments}

\bibliography{biblio}

\renewcommand{\thefigure}{S\arabic{figure}}
\setcounter{figure}{0}
\renewcommand{\theequation}{S\arabic{equation}}
\setcounter{equation}{0}

\section{Supplemental Material}

In this Supplemental Material, we present the details of the experiments, simulations, and calculations given in the main text. We discuss the full Schr\"odinger equation obtained in the paraxial approximation. We provide the details on the analytical solution of the Dirac equation, comparing it with a numerical solution of the same equation. Finally, we present the supplementary video files.

\subsection{Additional details on the experimental measurements}
In this section, we provide additional comments on the experimental scheme presented in Fig.~1 of the main text.

The coupling beams $\emph{\textbf{E}}_{2}$, $\emph{\textbf{E}}_{2}'$ and $\emph{\textbf{E}}_{2}''$ from the same tunable continuous-wave diode laser ECDL2 (wavelength around 795 nm) are coupled symmetrically with respect to \emph{z} axis by three PBSs and intersect at the center of the Rb vapor cell to establish the hexagonal interference pattern in the \emph{x-y} plane. The angle between each coupling beam is 2$\theta\approx$1.2$^\circ$ and the period of the formed hexagonal lattice is about a 25$\mu m$ [Fig. 1(g)]. 
The powers of the three Gaussian coupling beams (with the same diameter of $\sim 0.8$~mm) are the same: 15~mW. 
The resonant wavelength (corresponding to detuning $\Delta_2=0$) of the coupling field is $\lambda_2=794.975$~nm.
The 5cm long atomic vapor cell is wrapped with $\mu$-metal sheets to shield outside magnetic field and heated by a heat tape to 80$^\circ C$. The two probe beams $\emph{\textbf{E}}_{1}$ and $\emph{\textbf{E}}_{1}'$ from the same tunable continuous-wave diode laser ECDL1 are also symmetrically placed with respect to \emph{z} axis and intersect at the center of the cell with almost same angle as for the coupling beams to build the probe-field lattice along the transverse direction \emph{x}. 
The two cw Gaussian probe beams (with the same diameter of $\sim 1$~mm) are at $\sim 0.5$~mW. The resonant wavelength (corresponding to detuning $\Delta_1=0$) of the probe field is $\lambda_1=794.981$~nm.
Under the EIT condition, the honeycomb refractive index lattice can be effectively written inside the coherently-prepared multi-level atomic medium. 
The induced change of the refractive index (see below) is of the order of $9\times 10^{-3}$ for the real part and $1.5\times 10^{-4}$ for the imaginary part.
The decay rates between the states involved in EIT are determined by the the longitudinal and reversible transverse relaxation times of each state.
The two probe beams exiting the cell are interfered with a Gaussian-shaped reference beam from ECDL1, introduced into the optical path via a 50/50 beam splitter.

\subsection{Vortex dynamics in the full Schr\"odinger equation}
In the main text, we explained that the transverse profile of a beam in the paraxial approximation obeys the Schr\"odinger equation:
\begin{equation}
i\hbar \frac{{\partial \psi }}{{\partial t}} =  - \frac{{{\hbar ^2}}}{{2m}}\Delta \psi  + U\psi,
\label{schro}
\end{equation}
where the mass $m$ is determined by $m=\hbar k_0 n_0/c$ ($n_0$ is the background refraction index, $c$ is the speed of light), and the potential $U(x,y)$ is determined by the deviation of the refraction index from the background value: $U(x,y)=-\hbar c k_0^2(n^2-n_0^2)$.

At $t=0$, a wave packet is created in the potential minima corresponding to the $A$ sites. We then follow its evolution and analyze its phase by studying its interference with a reference beam. Both for the theory and for the experimental images, we extract the phase from the interference pattern by making a Fourier transform, keeping only one of the maxima, shifting it to $k=0$, and finally making an inverse Fourier transform.

\begin{figure}[h]
\includegraphics[width=1\linewidth]{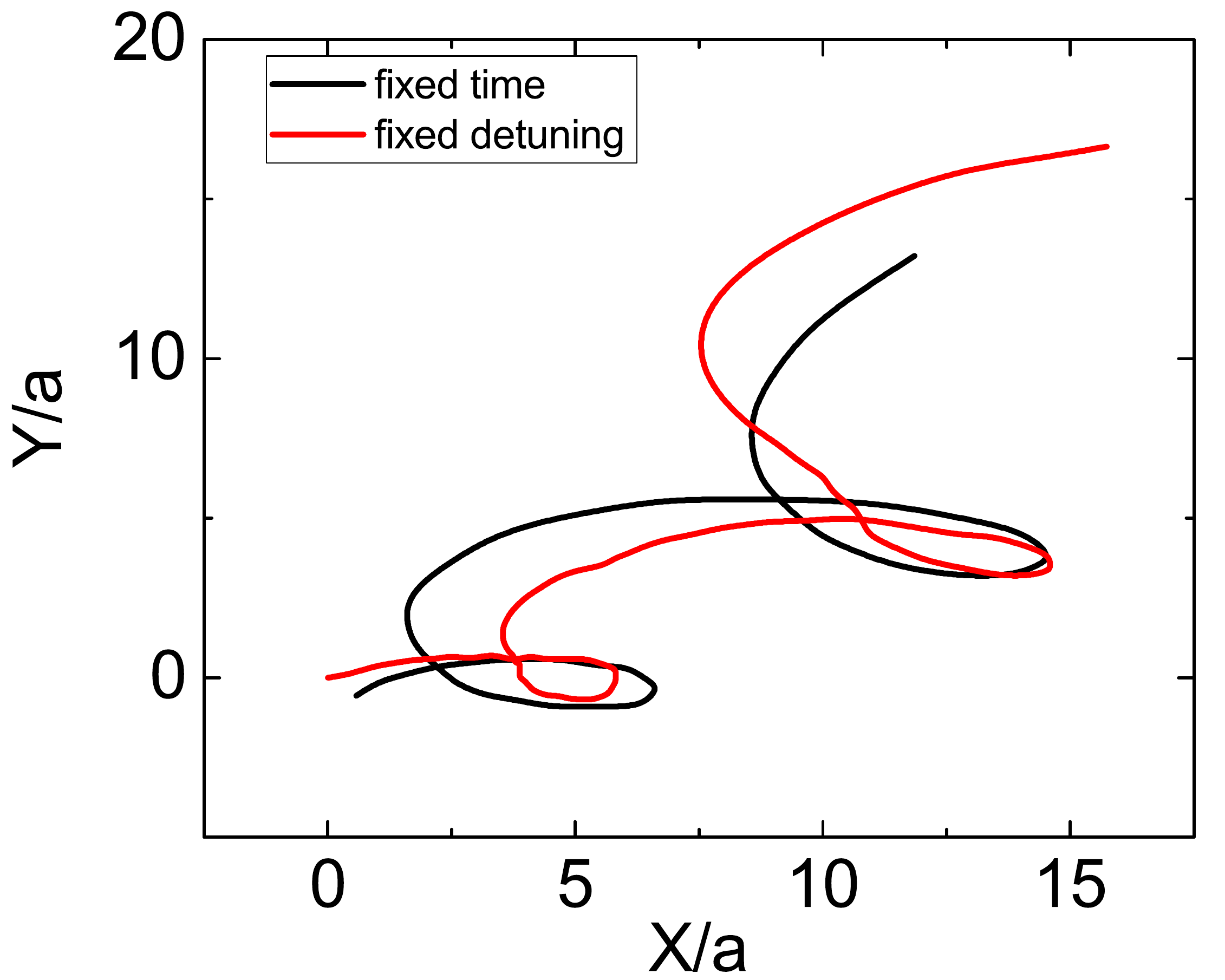}
\caption{\label{figS5} Vortex trajectory from full paraxial equation for fixed time (varying the detuning and thus the effective potential -- black line) and for fixed detuning (varying the propagation length and thus the duration of the evolution -- red line).}
\end{figure}

In the main text, we explained that due to the fact that the Dirac equation contains only a single parameter $c$, the study of the time evolution of a system with a fixed parameter $c$ is equivalent to the study of the evolution of multiple systems with different $c$ during a fixed time. Of course, this is only completely true at the level of the Dirac equation: for a full paraxial equation, changing the potential affects the spatial profile of the Bloch functions, which can lead to deviations from the equivalence of the two configurations. Here, we demonstrate that the qualitative behavior observed in the two cases is essentially the same. Figure~\ref{figS5} shows the trajectories of the vortex obtained for fixed time (black line) varying the effective potential height (which is equivalent to changing the detuning in experiment), and for fixed detuning (red line) varying the evolution duration (which would be equivalent to changing the length of the Rb cell in experiment). Both curves show qualitatively the same behavior, which means that both configurations are indeed equivalent. The differences in the two trajectories are due precisely to the fact that they cannot be completely obtained within the Dirac equation, and therefore do not completely retain the associated invariance.

The values of the detuning available in experiment allow the observation of effective times approximately from $0.2$ to $1.5 T$, where $T$ is the conversion period for a fixed wavepacket width of $w=50~\mu$m.

\subsection{Analytical solution for the vortex trajectory in the Dirac equation}
The 2D Dirac equation for massless particles can be written as
\begin{equation}
i\hbar\frac{\partial\psi}{\partial t}=\hbar c' \mathbf{k}\cdot\mathbf{\sigma} \psi,
\label{dirac1}
\end{equation}
where $\mathbf{k}$ is the wave vector (in 2D), and $\mathbf{\sigma}$ is a vector of Pauli matrices $\sigma_x$ and $\sigma_y$.
This equation can be obtained by linearization of the tight-binding Hamiltonian of graphene in the vicinity of the corner of the Brillouin zone (the so-called $K$ and $K'$ points).

\subsubsection{Initial wavepacket with $L=0$}
The approach used to find the analytical solution for the vortex trajectory in the Dirac approximation is discussed in the main text. It is based on a small-time approximation, which takes into account only the initial conversion $\psi_A\to\psi_B$ and neglects the backward conversion. The limit of validity of this approximation corresponds to the moment $t=T=w/c$, when $\psi_A$ vanishes. In this approximation, we are keeping a Gaussian shape for $\psi_A$ and a corresponding derivative shape for $\psi_B$. Of course, at any moment of time, in reality there is a backwards conversion from $\psi_B$ to $\psi_A$, and therefore the shape of $\psi_A$ is not Gaussian any more. But this backwards conversion is a second-order perturbation, and this is why it does not influence the overall dynamics of the vortex. Even when we are at one half of the full conversion cycle [Fig.~2, panels (d,e)], when this non-Gaussian perturbation becomes comparable with the intensity of the original Gaussian in $\psi_A$, it does not influence the trajectory of the vortex, because the overall intensity $|\psi_A|^2$ is at this moment much smaller than $|\psi_B|^2$, and therefore the vortex is necessarily located at the minimum of $|\psi_B|^2$. So, the analytical solution works well in spite of the approximations used.

The solution found in the main text is:
\begin{equation}
\left|x\left( t \right)\right| = \frac{{{w^2}\cot \omega t}}{{ct}},
\label{traj}
\end{equation}
In the main text, we compare this solution with the coordinates of the vortex center extracted from the experiment. Here, we compare it in Fig.~\ref{fig2} with the vortex trajectory extracted from the numerical simulation based on the direct solution of the Dirac equation Eq.~\ref{dirac1} (and not on the full Schr\"odinger equation), in order to check the validity of the underlying simplifications. The analytical solution is plotted as a red curve, while the numerical trajectory [obtained from the interference pattern shown in Fig.~2(c,f)] is plotted as black dots because of the finite step size in the numerical simulation. The good quality of the fit confirms the interpretation. We plot the figure in log-log scale to make visible the equally good agreement for all points (with relatively large and relatively small values).

\begin{figure}[h]
\includegraphics[scale=0.33]{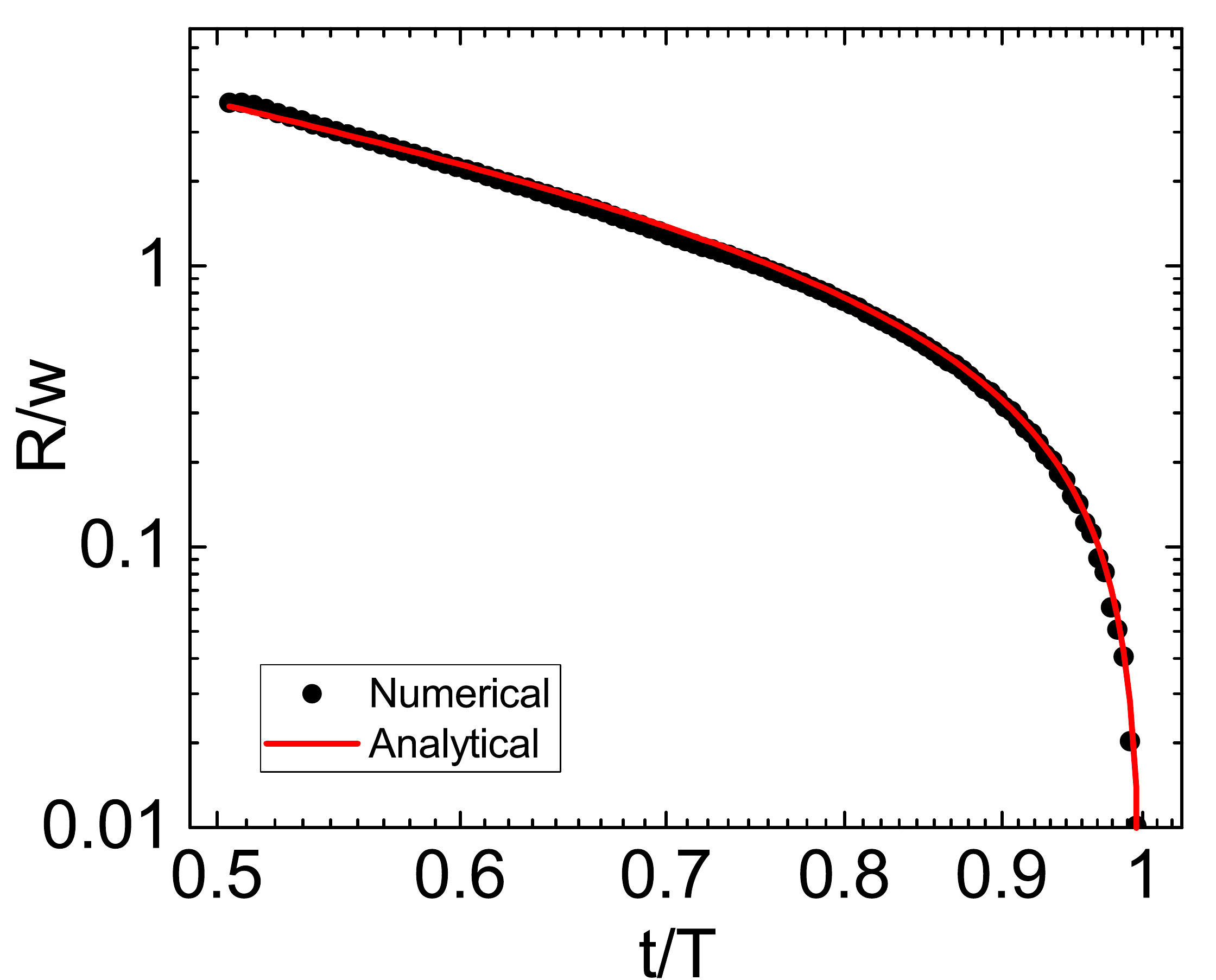}
\caption{\label{fig2} The trajectory of the vortex in the total wavfunction $\psi_A+\psi_B$: black dots -- extraction from the interference pattern in the numerical simulation, red line -- analytical solution.}
\end{figure}

\begin{figure}[h]
\includegraphics[width=1\linewidth]{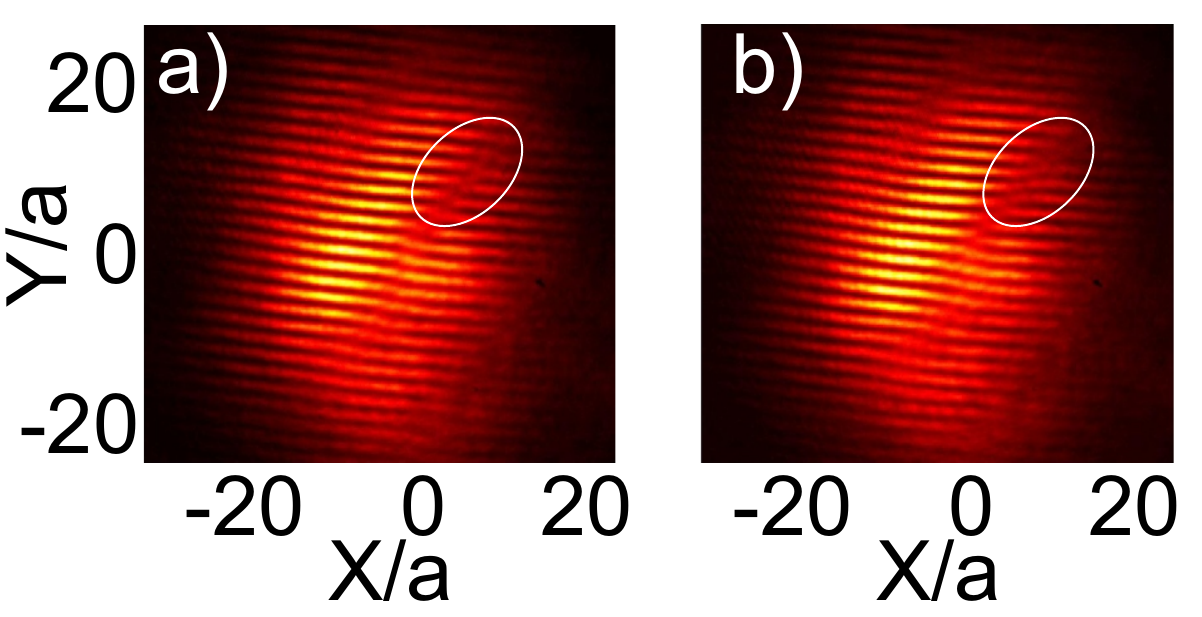}
\caption{\label{figS4} Experimental images showing spatial interference patterns, showing the vortex appearance: a) no vortex b) vortex (and anti-vortex at the pair creation point).}
\end{figure}

The analytical solution predicts that the vortex appears at infinity and approaches the center of the system with an infinitely high initial velocity. In numerical simulations and in experiment it is, of course, impossible to observe such behavior because of the finite precision. Indeed, the Gaussian function describing the wavepacket rapidly decays to very small values, smaller than the noise in experiment or than the numerical precision in simulations. It is in this region that the vortex actually appears. The quantization of the topological charge of the system implies that this vortex can only appear inside the system (not at infinity) as a part of a vortex-antivortex pair, and this is what is actually observed.  Figure~\ref{figS4} shows the experimental images with the interference patterns just before and after the vortex formation. In panel (a), there is already a shift of the interference, but no forklike dislocations are visible. In panel (b), the vortex-antivortex pair is shown after its separation. The anti-vortex remains where the pair has appeared, and the vortex moves towards the center of the system, as described by the analytical solution Eq.~\eqref{traj}. We note that the formation of the vortex-antivortex pair occurs at a large distance from the center of the system, where the intensity of the probe beam (not the reference beam used for interference) strongly drops down.

\subsubsection{Initial wavepacket with $L=\pm 1$}
If the original wavepacket contains a non-zero angular momentum, the overall outline of the theoretical consideration does not change in the Dirac approximation. The conversion of the component $\psi_A$ into $\psi_B$ changes the angular momentum by $1$: $l_B=l_A-1$, because of the shape of the Dirac Hamiltonian. If the injected wavepacket has a positive angular momentum $l_A=1$, then the other component will have $l_B=0$, and the overall interference pattern will show a vortex and an anti-vortex, canceling each other at $t=t_0$. In the other case, when $l_A=-1$, $l_B=-2$, and the overall pattern will show $L=-2$ at $t_0$.

\begin{figure}[h]
\includegraphics[scale=0.33]{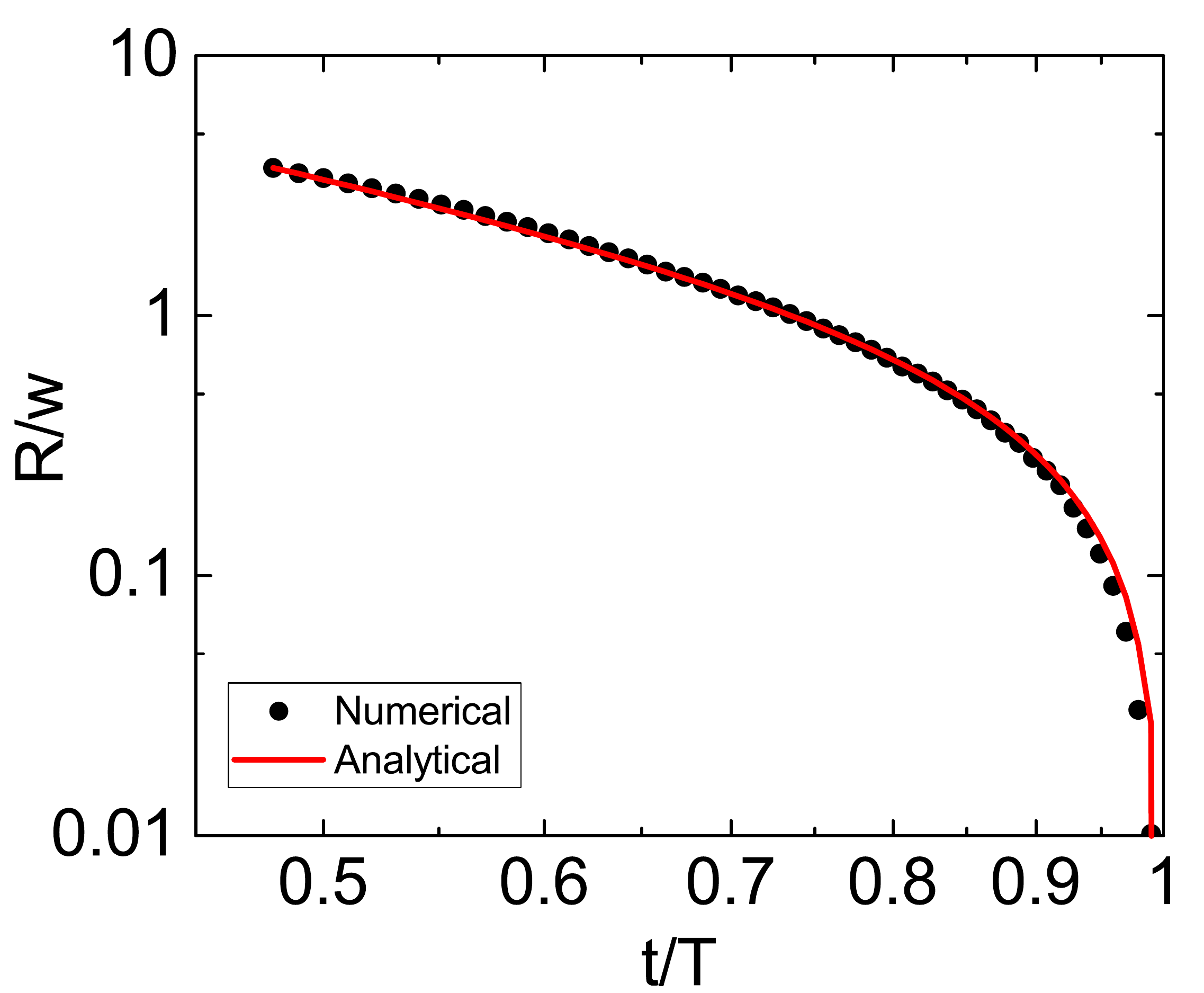}
\caption{\label{fig6} The trajectory of the vortex in the total wavfunction $\psi_A+\psi_B$ for $l_A=-1$, $l_B=-2$: black dots  -- extraction from the interference pattern in the numerical simulation, red line -- analytical solution.}
\end{figure}

As in the previous case, here it is also possible to find an analytical solution for the vortex trajectory from the condition $\psi_A+\psi_B=0$. Close to $t=0$, the solution remains the same, while for $t\to t_0$ ($x\approx 0$) there are second-order corrections to the trajectory due to the modified spatial dependence of the wavefunction $\psi_A$ in the presence of a vortex: $\psi_A(x)\propto x\exp(-x^2/2w^2)$:
\begin{equation}
x\left( t \right) = \frac{{{w^2}\cot \omega t}}{{ct}} + 2\frac{{{w^3}{{\cot }^2}\omega t}}{{{c^2}{t^2}}},
\end{equation}
The numerical (black dots) and analytical solutions (red line) for the trajectory of the extra vortex for $l_A=-1$ are shown in Fig.~\ref{fig6}. Again, a perfect fit is obtained, confirming the theoretical assumptions. An important difference with the previous case is that here the overall dynamics is much faster, because the effective width of the wavepacket is modified by the change of the spatial profile (higher average wave vector for a rotating wave packet), which increases $\omega$ by approximately a factor 2. This is not visible in dimensionless coordinates $t/T$, except via the decreased density of numerical points for a fixed time step.
The other vortex always remains at $x=0$, so there is no need to calculate its position (this is only valid within the framework of Dirac equation).

\subsection{2-beam vs 3-beam configuration}

In this section, we compare the possible experimental configurations with 2 or 3 beams exciting correspondingly 2 or 3 equivalent valleys, stressing that there are no fundamental differences between the two cases. The 2-beam configuration was used in the present work, while the 3-beam was used in Ref. \cite{Song2015}.

\begin{figure}[h]
\includegraphics[width=1\linewidth]{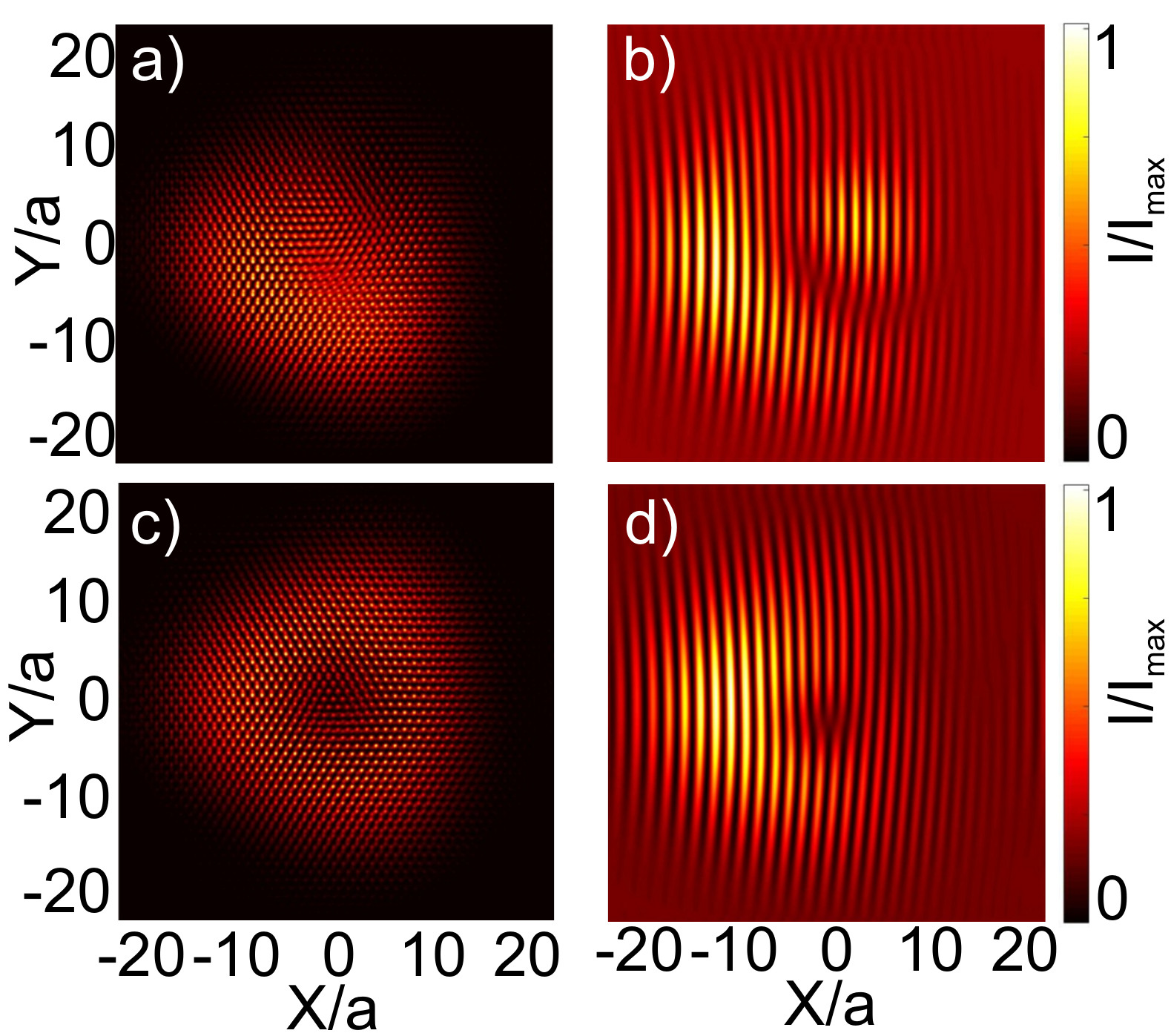}
\caption{\label{figS3} Left column: spatial images of the intensity distribution $I(x,y)$ for 2-beam (a) and 3-beam (c) configurations. Right column: interference pattern for a single extracted beam (valley): 2-beam (b) and 3-beam (d) configurations.}
\end{figure}

The interpretation of the experiments with an emergent Dirac equation or similar ones is essentially based on the decomposition of the wavefunction into a slowly-varying envelope and a plane wave, corresponding, for example, to a certain valley. The solution of the Dirac equation gives the behavior of the envelope function (the wave packet and a part of the Bloch wave in the spinor), which then has to be multiplied by the plane wave of the valley. The corresponding state thus includes an interference of the complex exponents corresponding to the wave vectors of the 3 equivalent valleys, giving rise to a complex interference pattern with a vortex inside each unit cell \cite{Bleu2018}. It is therefore quite difficult (but possible) to observe an extra vortex appearing in the envelope function by studying the intensity profile of the whole solution, as shown in Fig.~\ref{figS3}(c), calculated using the full Schr\"odinger equation Eq. \eqref{schro} starting with a Gaussian wavepacked composed of 3 beams (plane waves) of the K valley. However, the 3 different wave vectors of the 3 equivalent valleys forming the Bloch function naturally separate in space during the propagation of light after exiting the Rb cell, and it becomes possible to study just one of them, bearing only the information on the envelope function modulated by a single complex exponent with one definite wave vector [Fig.~\ref{figS3}(d)]. The same applies to the 2-beam configuration: while the 2 beams do not give the Bloch wave perfectly, this does not affect the evolution of the envelope function. Indeed, the distribution of intensity in Fig.~\ref{figS3}(a) corresponding to the 2-beam configuration is slightly different from Fig.~\ref{figS3}(c), but the phase pattern of the envelope function extracted from one of the 2 beams still shows a vortex in the center.  This confirms the generality of the results obtained in the present work.

We note, however, that if only a single beam corresponding to one of the three equivalent valleys is used, the spatial image of the conical refraction becomes too much perturbed. This affects the interference pattern, and the extraction of the vortex position becomes difficult or even impossible.

\subsection{Spatial dynamics and decay}

\begin{figure}[h]
\includegraphics[width=1\linewidth]{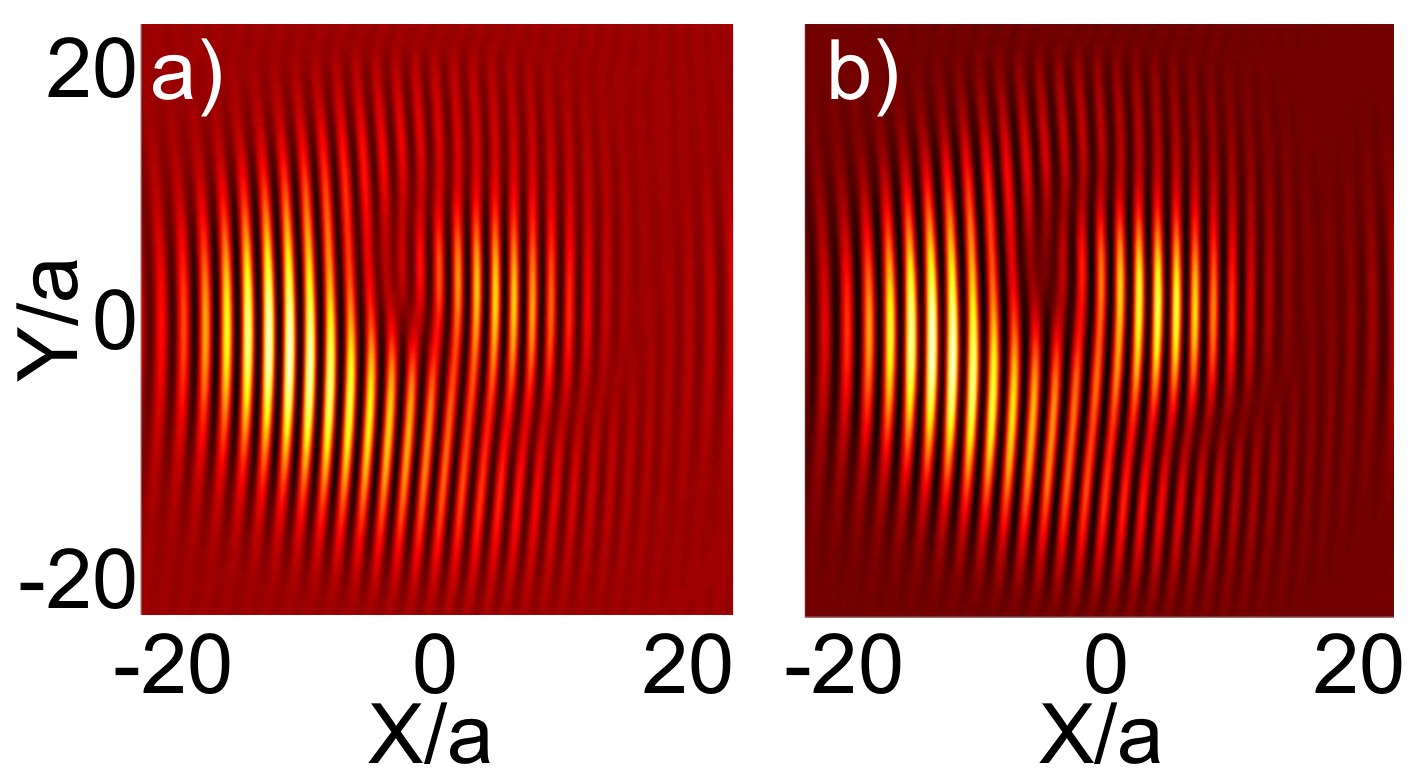}
\caption{\label{figS6} Calculated spatial images of the interference with a reference beam for (a) purely real potential (b) potential with a small imaginary part (3\% of the real part).}
\end{figure}

The EIT used for the creation of the photonic graphene lattice in our experiment induces not only the modulation of the real part of the refraction index, but also of its imaginary part. However, the definition of the effect of electromagnetically induced transparency suggests that the absorption should be low. The experimental estimates of the real and imaginary part of the induced refractive index change suggest that the ratio between the two is of the order of several percents. In all our simulations in the main text the imaginary part of the potential was neglected. Here we show that the results do not change qualitatively if a 3\% imaginary potential is taken into account. Figure~\ref{figS6} compares the calculated interference profiles for a purely real potential (a) and for a potential with a 3\% imaginary part (b). No qualitative differences are visible neither in the interference pattern, nor in the underlying intensity profile (but the overall intensity is of course decreasing). We note, however, that these results can change qualitatively if the absorption becomes higher. For a Gaussian wave packet, the transition occurs roughly at $\Gamma\approx c/w$, where $w$ is the width of the wave packet. Indeed, the imaginary part of the potential induces a particle decay and a broadening of the dispersion $\Gamma$. If the wavepacket is completely within the broadening, it cannot exhibit any dynamics, because the system becomes effectively overdamped at the corresponding scale. All experimental observations confirm that we are not in such a regime in our case.

\subsection{Supplementary video files}
The supplementary video files were generated from sequences of experimental interference images, obtained at different detunings, which corresponds to different moments of time, as discussed in the main text. The cores of the vortices (visible as phase dislocations) are marked with white crosses, which are obtained automatically by analyzing the phase maps.

\begin{enumerate}
\item \textsf{fig2interf.mp4} -- Experimental images, corresponding to Fig.~2 of the main text (no angular momentum in the initial wavepacket).
\item \textsf{fig2theory.mp4} -- Theoretical images, showing the curl of the phase of the wavefunction obtained from the full Schr\"odinger equation, together with the contour of the potential. The vortex is visible as a blue spot. This simulation corresponds to Fig.~2 of the main text (no angular momentum in the initial wavepacket). The parameters of the full numerical simulations were chosen different from those of the experiment in order to enhance the Magnus force, leading to a prolate cycloidal trajectory with a visible retrograde motion.
\item \textsf{fig3interf.mp4} -- Experimental images, corresponding to Fig.~3 of the main text (initial wavepacket with $L=+1$, two vortices of opposite signs).
\item \textsf{fig4interf.mp4} -- Experimental images, corresponding to Fig.~4 of the main text (initial wavepacket with $L=-1$, two vortices of the same sign).
\end{enumerate}

\end{document}